\documentclass{segabs}

\usepackage{algorithm}
\usepackage[noend]{algpseudocode}

\begin{document}

\title{Full waveform inversion with random shot selection using adaptive gradient descent}
\vspace{0.2in}
\renewcommand{\thefootnote}{\fnsymbol{footnote}} 

\author{Bharath Shekar$^{{\ast}}$  \\
		$^{1}$Department of Earth Sciences, Indian Institute of Technology Bombay, Mumbai, India. \\
        $^{{\ast}}$  bshekar@iitb.ac.in}

\righthead{Random shot FWI with adaptive gradient descent}

\maketitle

\begin{abstract}

Full waveform inversion (FWI) is a powerful yet computationally expensive technique that can yield subsurface models at high resolution.  Randomly selected shots (``mini-batches") can be used to approximate the misfit and the gradient of FWI, thereby reducing its computational cost. Here, we present a methodology to perform mini-batch FWI  using the Adam algorithm, an adaptive optimization scheme based on stochastic gradient descent. It provides for stable model updates by smoothing the gradient  across iterations and can also account for the curvature of the optimization landscape.  We describe empirical criteria to choose the hyperparameters of the Adam algorithm and the optimal mini-batch size. The performance of the outlined scheme is illustrated on synthetic data from the Marmousi model.

\end{abstract}

\section*{Introduction}

Full waveform inversion (FWI)  is  a computationally intensive technique partly due to its requirement of repeated wave simulations \citep{FWI_overview2, FWI_overview}. Source encoding and random shot selection are two commonly employed starategies to decrease the computational cost of FWI. Source encoding can be implemented either during acquisition \citep{encoding_acquisition} by simultaneously injecting multiple sources with varying source signatures or during migration \citep{romero, perrone_encoding, TU} and FWI \citep{krebs, leeuwen, encoding_multiparam, restarted_lbfgs}  by convolving sequentially fired sources with randomly generated encoding sequences. The encoded sources act as a single effective spatially extended source, thereby reducing the number of per-source wavefield iterations.  

Random shot selection involves approximating the misfit and the gradient by a randomly chosen subset of shots at each iteration \citep{toto, FWI_noencoding, slbfgs, SGD_grad_smoothing}. The subset of selected shots constitutes a ``mini-batch" \citep{SGD_grad_smoothing}, borrowing from machine learning terminology. As prescribed by the theory of stochastic optimization \citep{robbins_monro}, the gradients are averaged over a certain iteration history to build the search direction.  Although stochastic optimization algorithms exhibit rapid reduction in the misfit function in the initial iterations, gradient averaging leads to slower convergence in the later iterations. \cite{SGD_grad_smoothing} hasten the convergence by preconditioning the gradient by structure oriented filters \citep{Hale} computed using migrated images. \cite{FWI_noencoding}  prescribe a gradual increase in the mini-batch size to benefit from the fast convergence associated with  stochastic  and conventional optimization algorithms in the initial and later iterations, respectively.  They calculate the model updates using an l-BFGS \citep{nocedalwright} based  optimization scheme introduced by \cite{friedlander}.  However, the l-BFGS method needs to be suitably modfied for application in the stochastic setting. The proposed modifications involve uncoupling the curvature and gradient estimates by updating the  inverse Hessian on a different schedule than the gradients \citep{byrd} or by updating the inverse Hessian using gradients from same set of sources in successive iterations \citep{slbfgs}.   \cite{subsampled_TN} introduce a stochastic truncated Newton algorithm, wherein the Hessian-vector product is formed using a non uniformly sampled set of  sources and the second order adjoint-state method  \citep{Hessain_fitchner, Hessian_second}. 

The deep learning era has spawned the development of a variety of algorithmic extensions to  stochastic gradient descent methods \citep{deeplearningbook}. \cite{richardson} and \cite{RNNFWI} show that acoustic wave propagation can be simulated using a recurrent neural network (RNN), leading to the formulation of FWI as that of training the RNN. They compare several optimization algorithms in training the RNN   and  find that Adam \citep{Adam}, a stochastic gradient based algorithm with adaptive moment estimation converges rapidly and yields stable results. Here, we apply  the Adam optimization algorithm to FWI with random shot selection, albeit {\it{without}} the need to simulate wave propagation  as an RNN. We first review the methodology of stochastic FWI with mini-batches, and then  discuss certain empirical criteria for the selection of hyperparameters with the aid of numerical examples.  Finally, we illustrate the performance of the algorithm on synthetic data generated from the Marmousi model. 

\section*{Stochastic full waveform inversion with mini-batches}

The data-misfit function for full waveform inversion with random shot selection can be defined as :
\begin{equation}
\mathcal{J}_{B}(\mathbf{\theta}) = \frac{1}{2B} \sum_{i \in \Omega_{B}}  \sum_{j=1}^{N_{r}}  \int_{0}^{T}  ||d^{\mathrm{obs}}(\mathbf{x}^{i}_{s}, \mathbf{x}^{j}_{r}, t) - d^{\mathrm{sim}}(x^{i}_{s}, x^{j}_{r}, t, \mathbf{\theta})||^{2}_2 \, \mathrm{d}t, 
\label{eq1}
\end{equation}
where  $\mathbf{\theta}$ represents a suitably discretized subsurface model,  $d^{\mathrm{obs}}$ and $d^{\mathrm{sim}}$ denote the observed and simulated data in the time domain and $T$ denotes the total recording time.  The  coordinate of the $i^{\mathrm{th}}$ source in a randomly chosen subset of sources $\Omega_{B}$ is denoted by $\mathbf{x}^{i}_{s}$.  The number of sources  in the subset  is denoted by $B$, and it  may be chosen to be less than or equal to the total number of sources, denoted by $N_{s}$.  The coordinates of the $j^{\mathrm{th}}$ receiver is denoted by $\mathbf{x}^{j}_{r}$,  and a fixed receiver spread  of $N_{r}$ receivers is assumed for all the shots.  A fixed receiver spread is  not a necessary condition and it is used here for the sake of convenience. Additionally, we  assume a constant density acoustic wave equation to model and invert seismic data. Henceforth, the model parameter $\mathbf{\theta}$ will refer to the P-wave velocity model.

Full waveform inversion with mini-batches proceeds by successively updating the model  along a search direction:
\begin{equation}
\mathbf{\theta}_{k+1} = \mathbf{\theta}_{k} + \alpha_{k} \, \mathbf{s}_{k},
\label{eq2}
\end{equation}
where  $\alpha_{k}$ is a suitable step-length and $\mathbf{s}_{k}$ is the search direction. The subscripts in equation~\ref{eq2} denote the iteration number. In the stochastic gradient descent scheme, the search direction is chosen as the negative of the gradient, i.e., $\mathbf{s}_{k} = - \partial \mathcal{J}_{B} / \partial \mathbf{\theta}_{k}$. The subset of shots chosen to form the misfit function $\mathcal{J}_{B}$ and the gradient varies with each iteration.  Hence, the gradient  at each iteration is a noisy estimate of the true gradient and needs to be smoothed across iterations \citep{FWI_noencoding}.  The gradient may also be suitably preconditioned, for instance, by structure oriented filters \citep{Hale} computed using migrated images \citep{SGD_grad_smoothing}.   Alternatively, second-order optimization methods suitably modified for the stochastic optimisation setting can also be used to scale the gradient by an estimate of the inverse Hessian \citep{FWI_noencoding, slbfgs}.  

Here, we use the Adam \citep{Adam} optimization algorithm to compute the updates to the velocity model.  The algorithm builds the search direction by computing the  first and second-order moments of gradient vectors weighted exponentially over the history of iterations to stabilize the direction of update and  correct for the curvature of the loss function, respectively.  


\section*{Numerical Examples}

\begin{figure} [t]
\centering
{
\subfigure[]{
\includegraphics[width=\columnwidth]{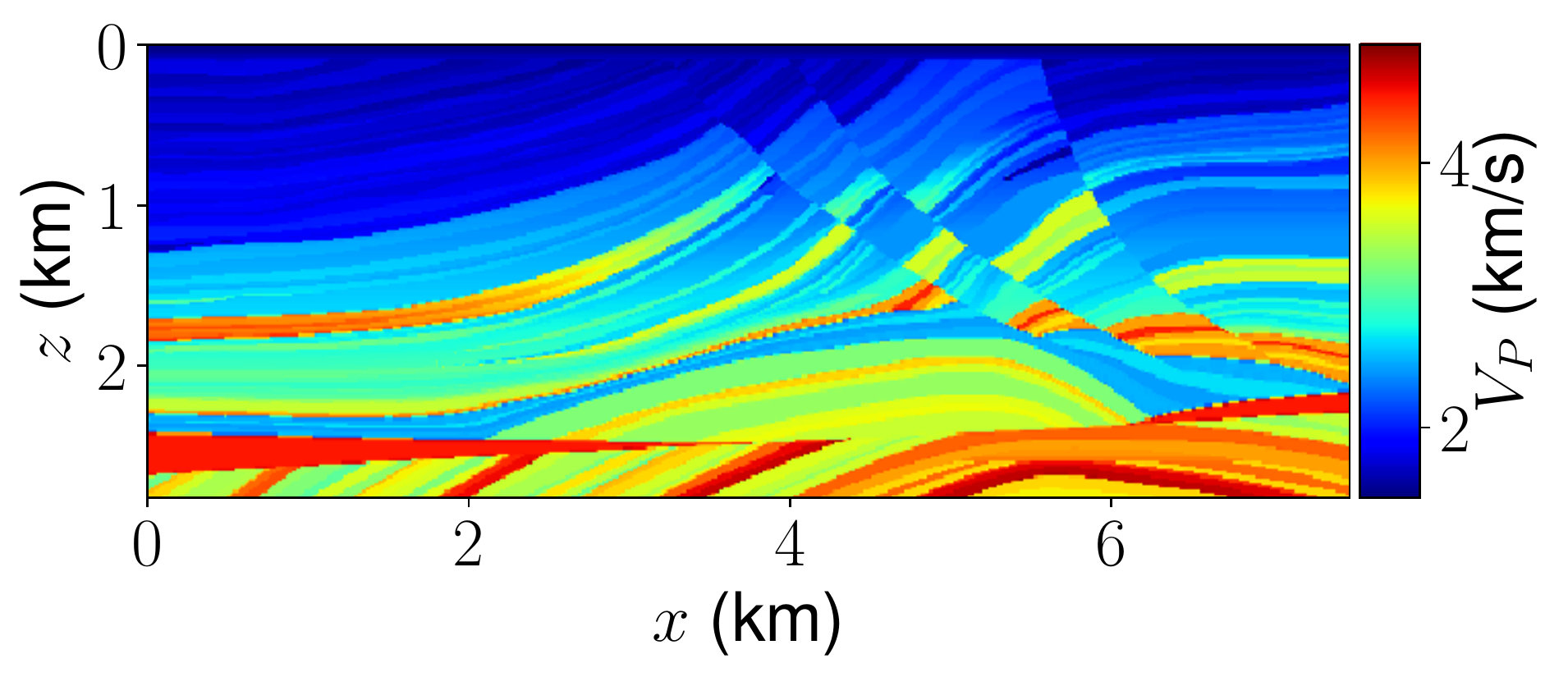}
\label{fig1-a}
}
\subfigure[]{
\includegraphics[width=\columnwidth]{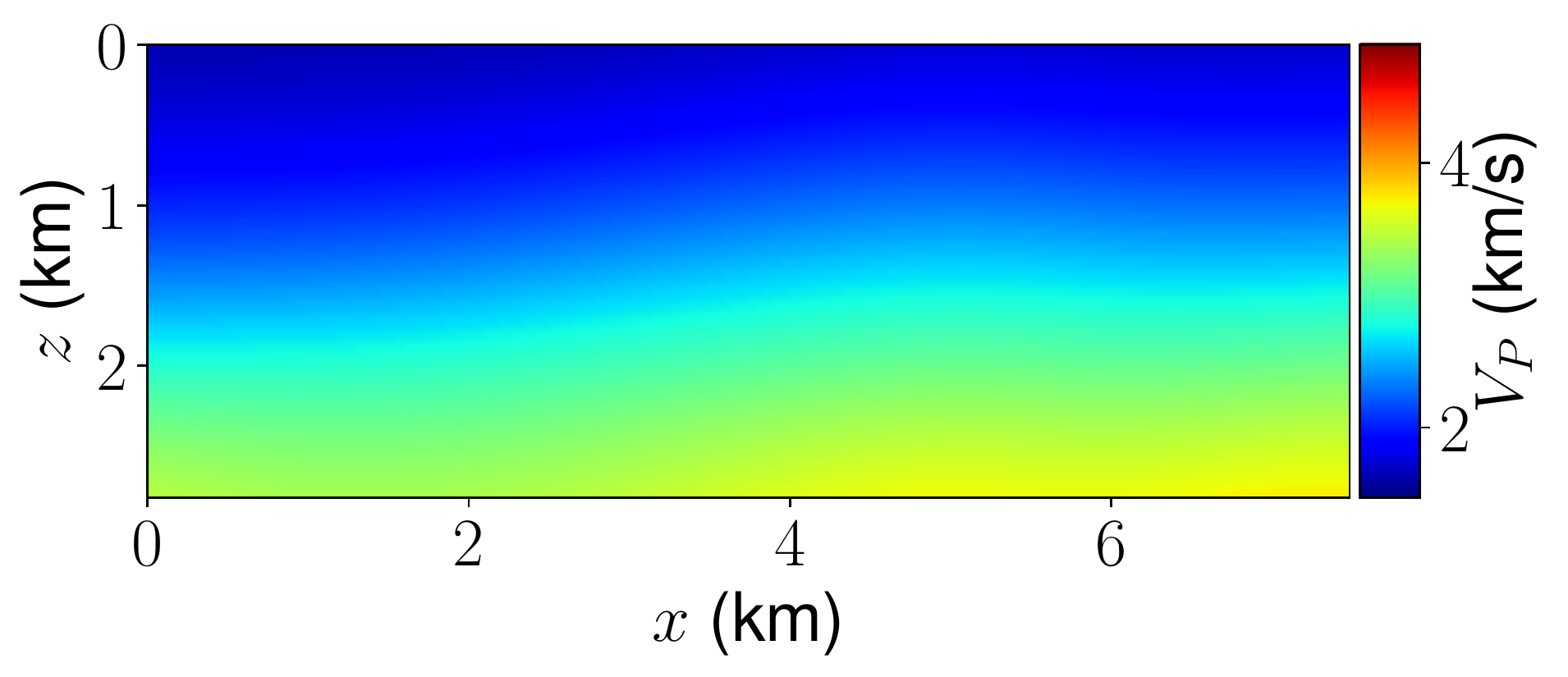}
\label{fig1-b}
}
}
\vspace{-0.2in}
\caption
{ P-wave velcoity models modified from the Marmousi model: (a) True and (b) initial velocity model.  }
\label{fig1}
\end{figure}

We illustrate stochastic optimisation strategies on a constant-density acoustic model modified from the  Marmousi model \citep{marmousi}. Figure~\ref{fig1} displays the true and initial P-wave velocity models used in the experiments.  The data are modeled with 32 evenly spaced sources each with the source signature as a Ricker wavelet with a peak frequency of 10 Hz.  The receivers  placed $30$ m below the surface with an interval of $15$ m in the $x$ direction  span the lateral extent of the model.  The receiver spread is fixed for all the sources, although it is not a necessary condition for FWI with random shot selection.  Acoustic wave propagation is simulated with the finite-difference method with absorbing boundaries (including the top)  using ``Devito", a domain-specific language (DSL) and code generation framework  \citep{devito-compiler, devito-api}.  The gradients are computed using the adjoint-state method \citep{plessix} implemented within the Devito package and the updates to the velocity model according to the Adam optimization algorithm \citep{Adam} are calculated using the Tensorflow package \citep{tensorflow2015-whitepaper}.  

\subsection*{Selection of hyperparameters}

The hyperparameters in the Adam optimization algorithm   namely the  learning rate $\alpha$, batch size,  $\beta_{1}$, $\beta_{2}$,  need to be set to values suitable for  FWI.  The symbols for the various hyperparameters are consistent with the notation used in \cite{Adam}, to aid easy reference. The parameters $\beta_{1}$ and $\beta_{2}$ control the exponential weights in the smoothed estimates of  first- and second-order moments  of the gradient vector computed over past iterations. The smoothing weights approximately decrease by a factor of $e$ in $1/(1-\beta_{1})$ iterations. Thus, the suggested value  $\beta_{1}=0.9$  and $\beta_{2}=0.999$ \citep{Adam} lead to an effective smoothing window of approximately 10 and 1000 iterations, respectively. A larger window is required for the second-order moment estimate in sparse settings \citep{Adam}. Most FWI applications do not involve a large number of iterations, furthermore, the FWI gradient is typically not sparse.  Therefore, we set $\beta_{1}, \, \beta_{2} = 0.9$, effectively averaging over 10 iterations to estimate the first- and second-order gradient moments. We found that the results were fairly robust to a range of values for $\beta_{1}$ and $\beta_{2}$ between 0.8 and 0.95.  The  parameter $\epsilon$ stabilizes the division in the model update \citep{Adam} and it is allowed to remain at the default value of $10^{-8}$.
 
\subsection*{Learning rate }

\begin{figure} [h]
\centering
{
\includegraphics[width=\columnwidth]{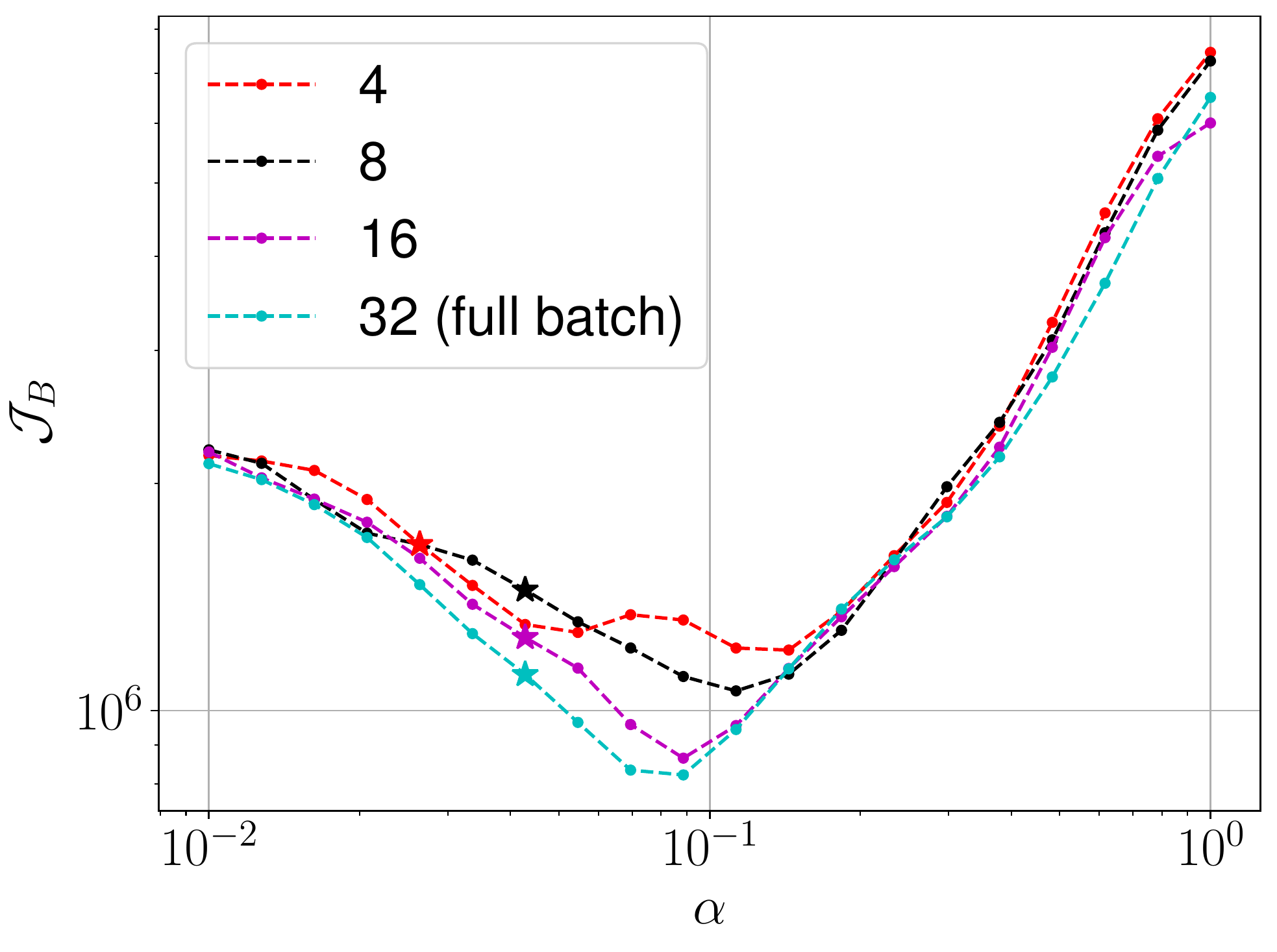}
}
\vspace{-0.2in}
\caption
{ Data-misfit values for varying mini-batch sizes as a function of exponentially increasing learning rate plotted in the log-log scale. The selected learning rate values are marked by stars.  }
\label{fig2}
\end{figure}

The performance of the Adam optimization algorithm is sensitive to the value of the learning rate $\alpha$. The learning rate is usually found by performing a grid search with cross-validation over a range of possible values \citep{deeplearningbook, richardson}.  We choose to optimize with a constant learning rate,  although  learning rate decay  may be used in stochastic gradient descent methods to improve convergence \citep{deeplearningbook}.  \cite{RNNFWI} adapt the value of the learning rate  by requiring the velocity model updates to be of the order of 10 m/s, since the magnitude of difference between the initial and true velocity models may be of the order 0 to 1000 m/s.  Hence, they search for  optimal learning rate in the interval [10, 100]. Here, the  interval for  the optimal learning rate is fixed between 0.01 and 1.0, since we use km/s for the units of the velocity models.  We follow the strategy described in \cite{learn_rate_finder} to choose the learning rate for algorithms based on stochastic gradient descent.  Although empirical, the criterion presented by \cite{learn_rate_finder}  has proven effective in finding optimal values of learning rates  in a variety of deep learning applications \citep{fastai}. It requires training the model for a few iterations with  variable and exponentially increasing learning rate values.  Figure~\ref{fig2} displays the  data-misfit value (equation~\ref{eq1}) for varying mini-batch sizes as a function of variable learning rate over 20 iterations of the optimization algorithm.  The mini-batches are formed by  sampling randomly from the 32 shots with replacement.   The heuristic for  the optimal learning rate \citep{learn_rate_finder, fastai} involves choosing a value smaller than that at the minimum and wherein a significant rate of change of data-misfit value is observed.   For the mini-batch size of 4, the hence chosen learning rate  is 0.02, and the value is 0.04 for mini-batch sizes 8, 16, and 32.  

\subsection*{Mini-batch size}

 \cite{friedlander} and \cite{FWI_noencoding} consider the error in the expected value of gradient computed from mini-batches to that from the full dataset.  They find that the error in gradient  is smaller for the case of random sampling without replacement than with replacement. Here, we follow the common practice in machine learning  \citep{deeplearningbook} and form mini-batches with uniform random sampling without replacement. All the shots are made available for sampling after an ``epoch", {\textit{i.e.}}, a single pass over all the 32 sources.  The computation of gradient  by the adjoint-state method lends itself to  data parallelism over shots. Hence, a single iteration over a batch may assumed to  take approximately equal computational time regardless of the mini-batch size, provided that sufficient number of threads are available. Furthermore,  an iteration over an epoch can be assumed to be proportional to the overall computational cost, defined empirically as the product of runtime and number of threads.   
 
\begin{figure} [t]
\centering
{
\subfigure[]{
\includegraphics[width=\columnwidth]{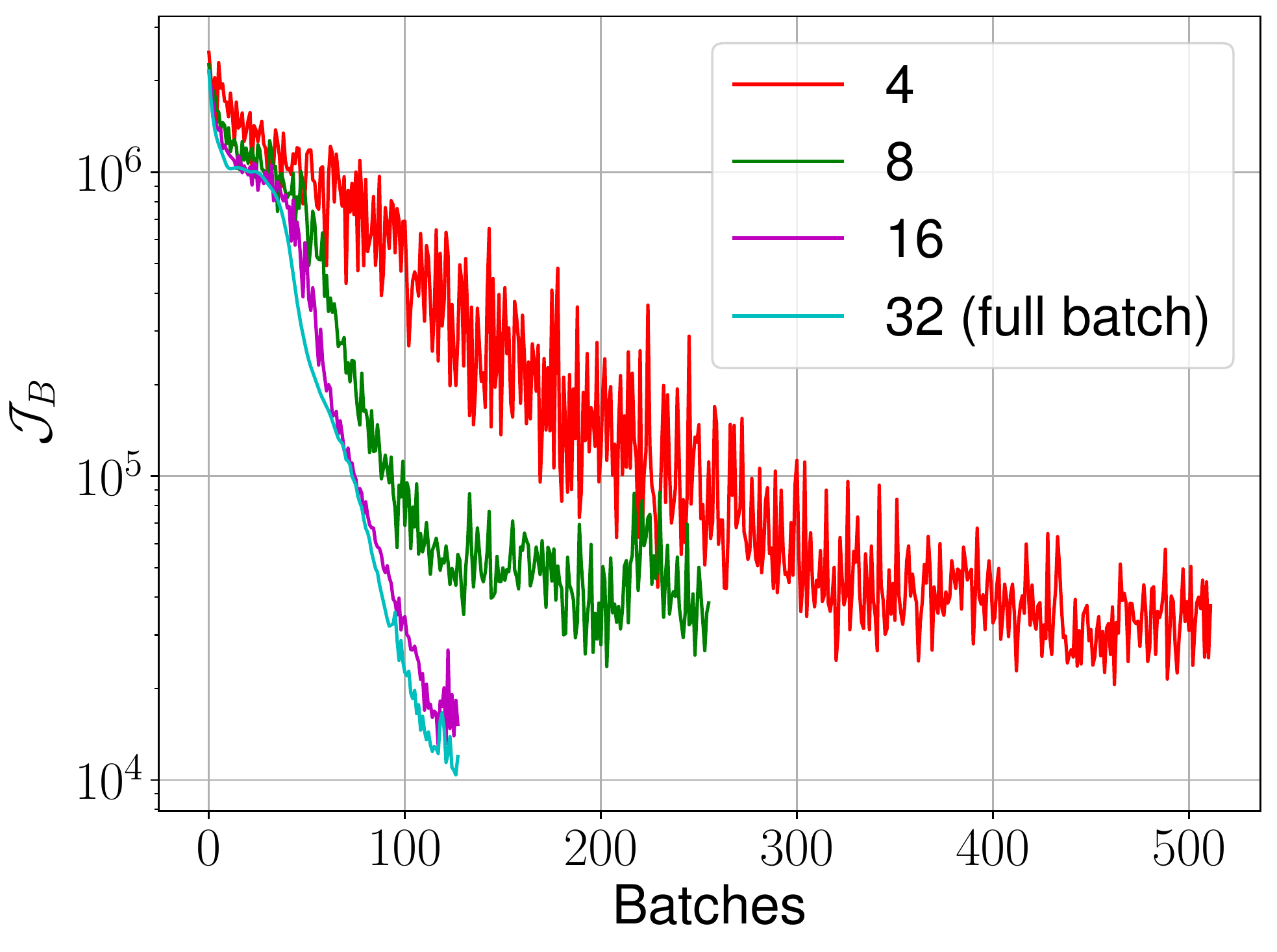}
\label{fig3-a}
}
\subfigure[]{
\includegraphics[width=\columnwidth]{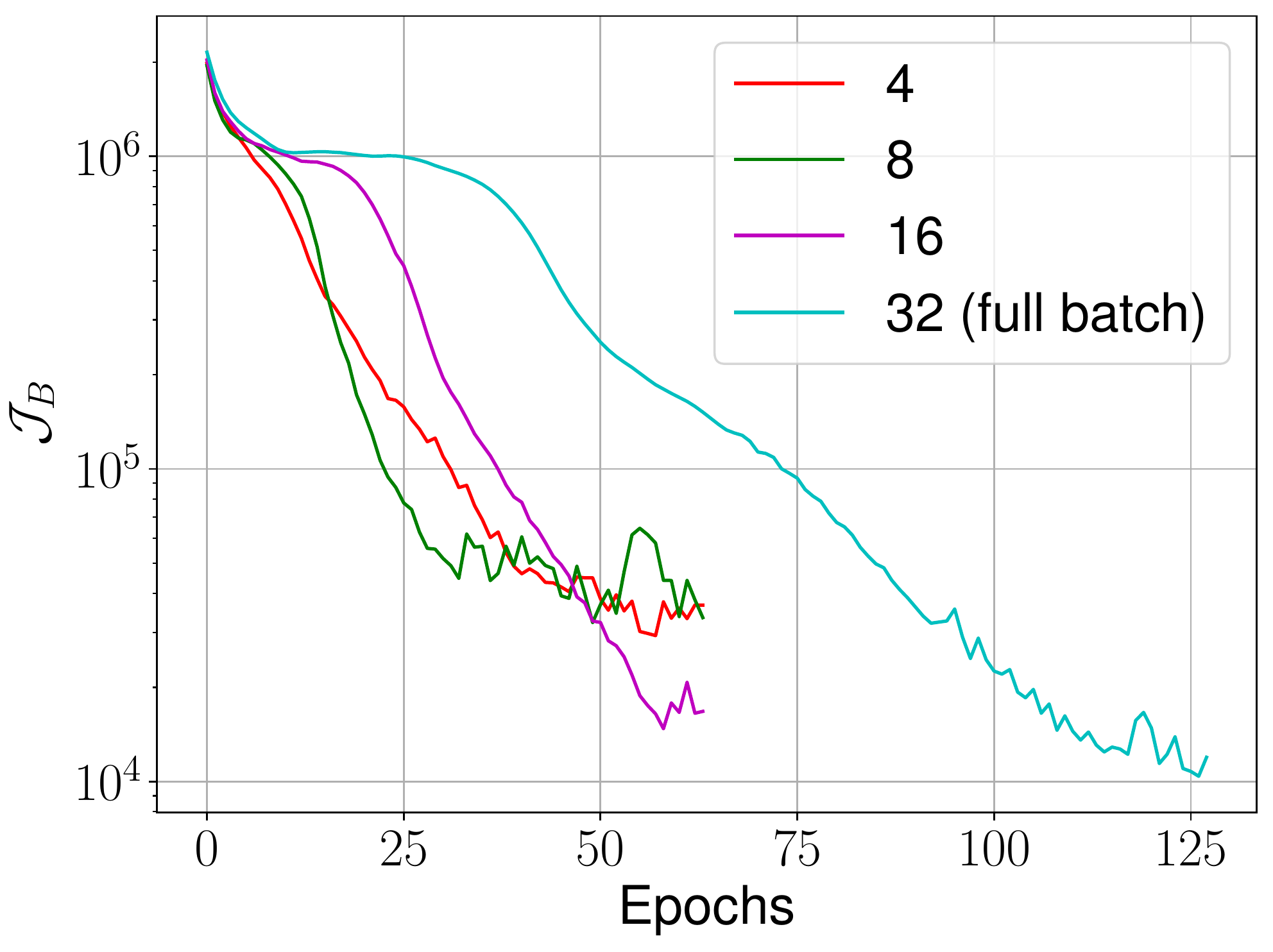}
\label{fig3-b}
}
}
\vspace{-0.2in}
\caption
{ Data misfit plotted in the logarithmic scale for varying mini-batch sizes as a function of the number of (a) batches and (b) epochs.}
\label{fig3}
\end{figure}

Figure~\ref{fig3} plots the data misfit in the logarithmic scale plotted as a function of the number of batches and epochs.  Inversion with smaller mini-batch size requires a larger number of batches for the data misfit value to saturate (Figure~\ref{fig3}a) indicating longer computational times. In Figure~\ref{fig3}b, the  misfit value for an epoch is found by summing across its constituent batches.  The reduction in data misfit is faster for smaller mini-batches as a function of epochs, indicating that FWI is computationally cheaper for smaller sized mini-batches. Interestingly,  the  data misfit for the mini-batch with 8 shots  saturates quicker (as measured by the number of epochs) than the mini-batch with 4 shots.  This behavior can be explained by the fact that the optimal learning rate value is 0.02 for the mini-batch size of 4, and the value is 0.04 for mini-batch sizes 8, 16, and 32.  Hence, the model updates are larger and the gradient is less noisier for the mini-batch with 8 shots than the one with 4 shots.  However, similar  computational gains are not observed for the mini-batch with 16 shots and the full batch with 32 shots since the learning rate is equal to 0.04 for both the instances.   Indeed, \cite{empirical_batch} argue that the tradeoff between computational time and cost in training deep neural networks  to achieve a specified level of performance takes the form of a Pareto frontier. Thus,  a mini-batch size of 8 is an optimal choice for the numerical experiment considered here.

\begin{figure} [t]
\centering
\vspace*{-0.1in}
{
\subfigure[]{
\includegraphics[width=0.95\columnwidth]{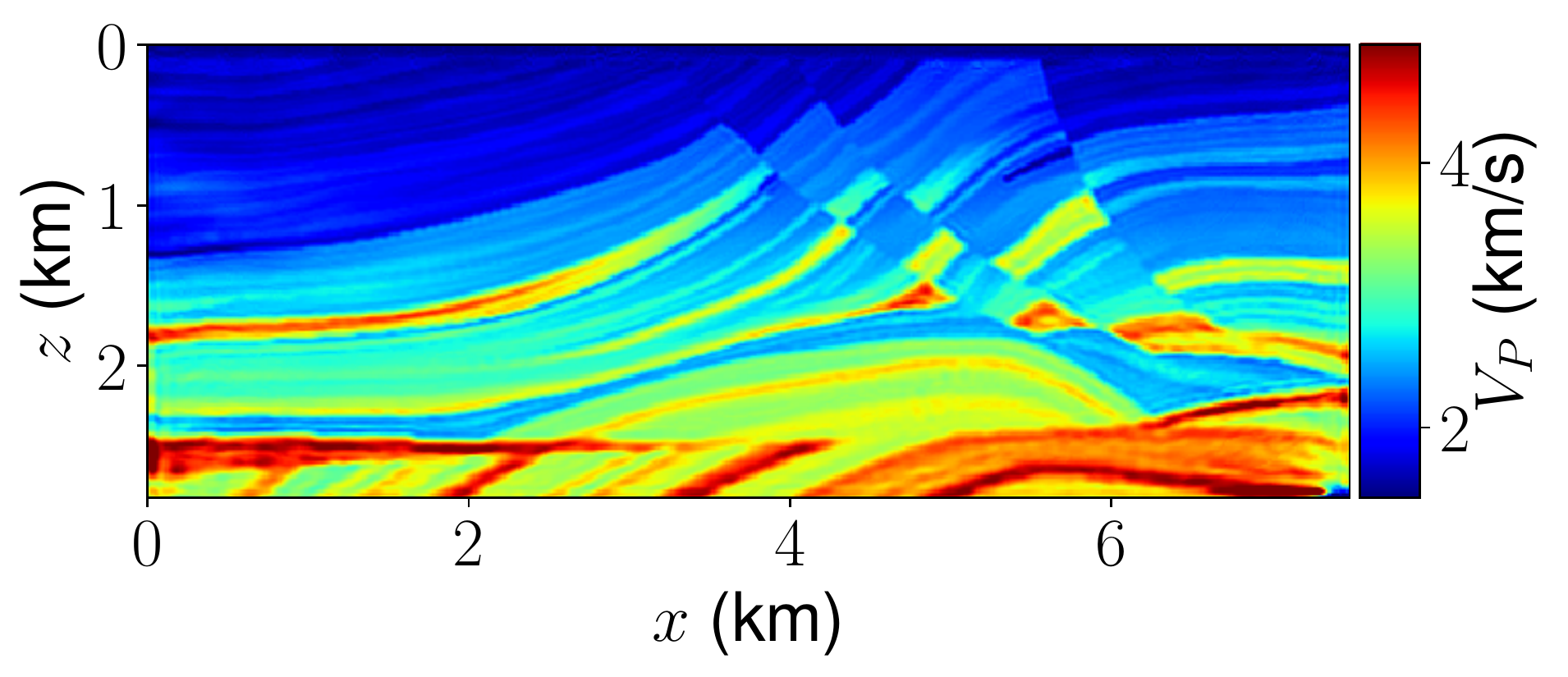}
\label{fig4-a}
}
\vspace*{-0.1in}
\subfigure[]{
\includegraphics[width=0.95\columnwidth]{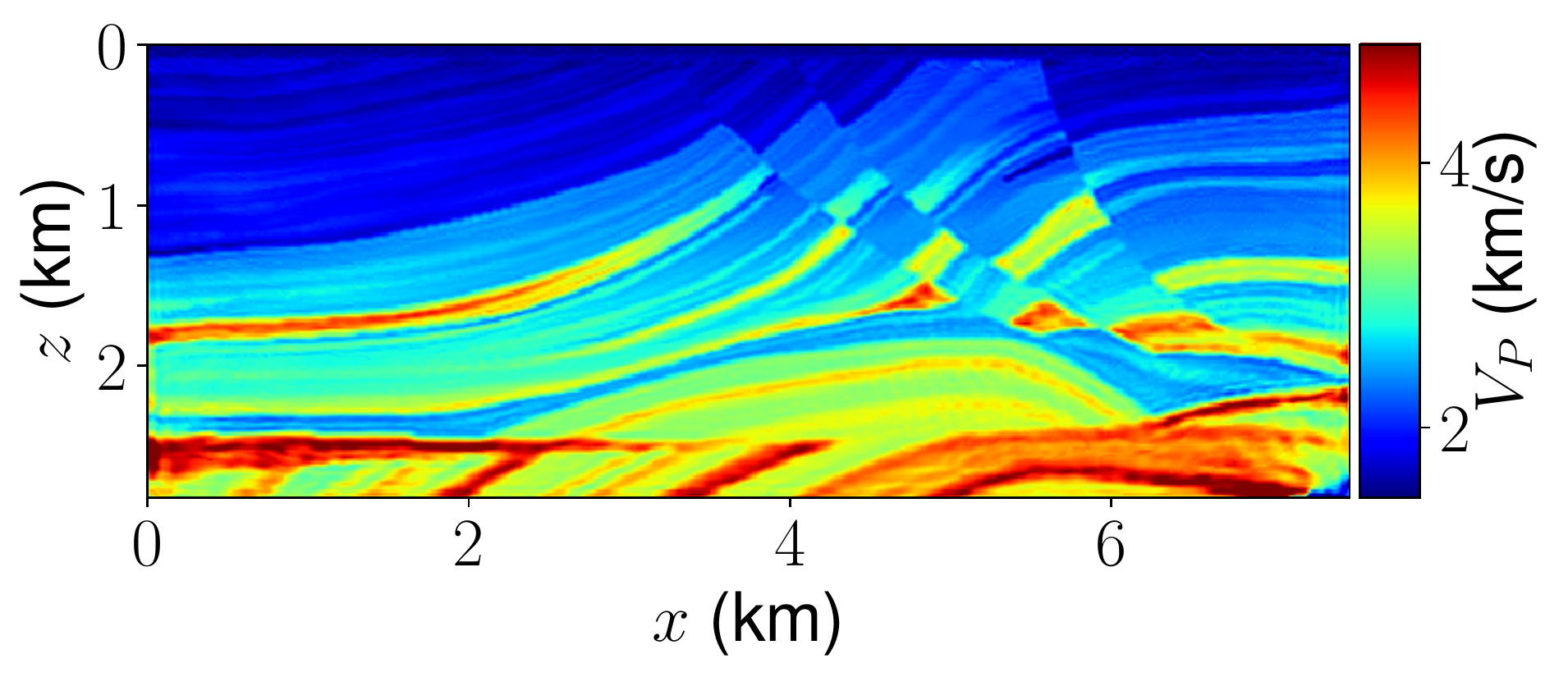}
\label{fig4-b}
}
}
\vspace*{-0.1in}
{
\subfigure[]{
\includegraphics[width=0.95\columnwidth]{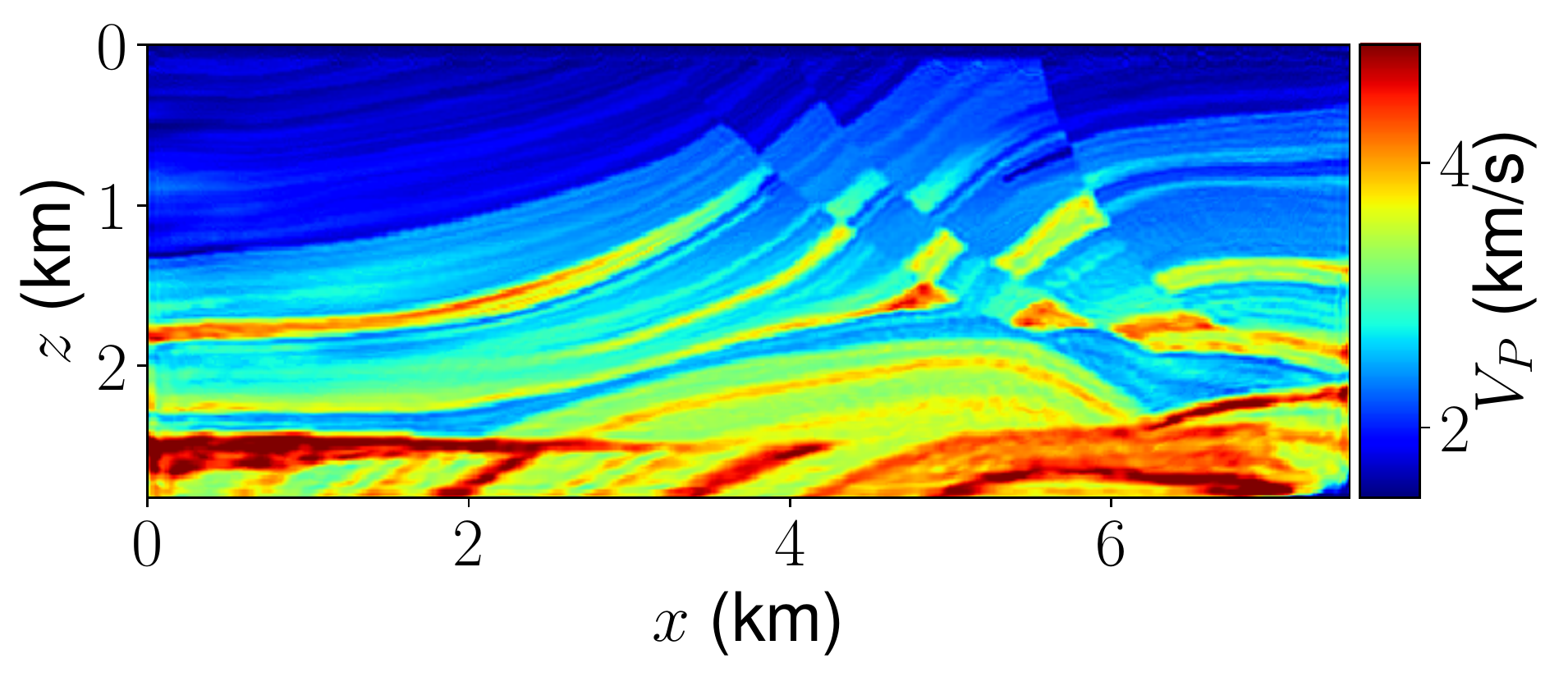}
\label{fig4-c}
}
\vspace*{-0.1in}
\subfigure[]{
\includegraphics[width=0.95\columnwidth]{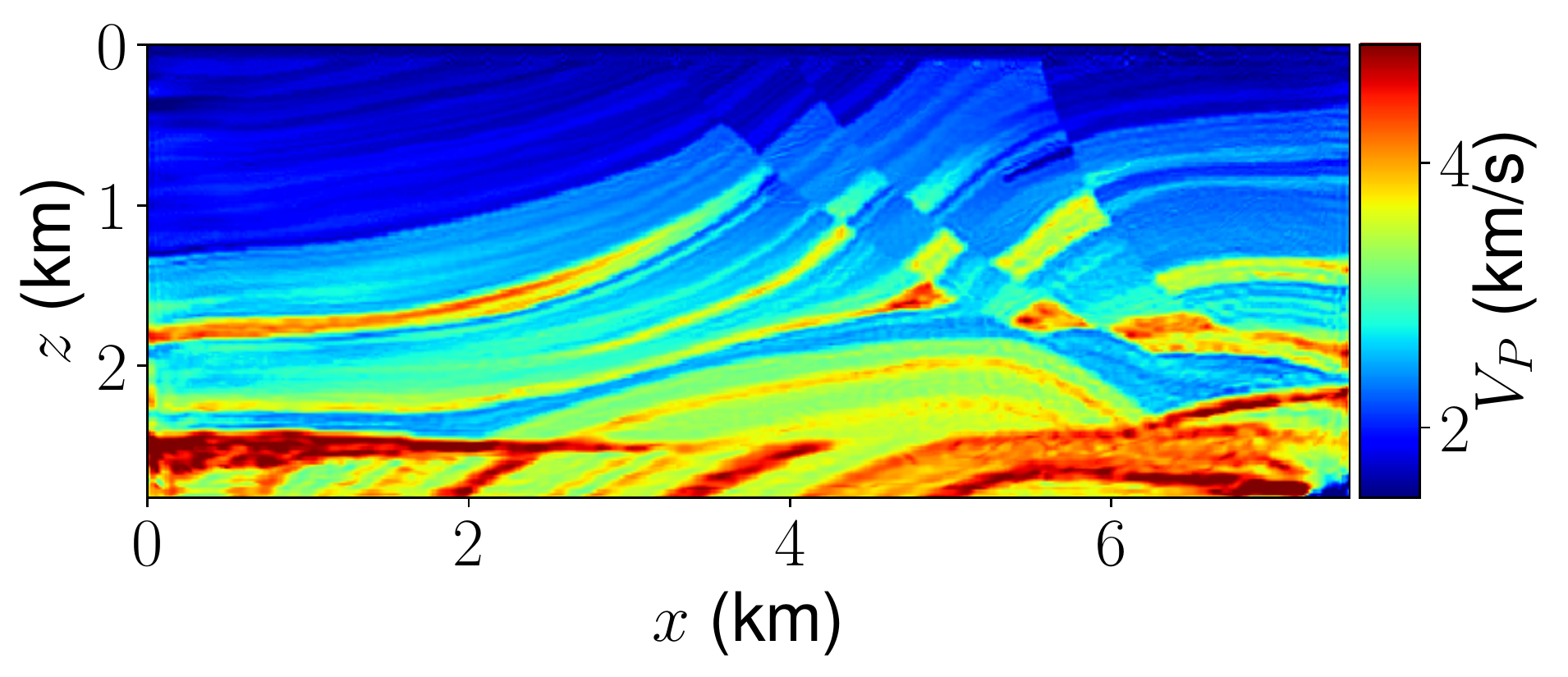}
\label{fig4-d}
}
}
\vspace*{-0.1in}
\caption
{Velocity model retrieved after 64 epochs and a mini-batch size of (a) 4,  (b) 8,  and (c) 16. The velocity model in (d) is retrieved after 128 epochs with the full-batch of data (32 shots). }
\label{fig4}
\end{figure}

\begin{figure}
\centering
\vspace*{-0.2in}
\noindent\makebox[\columnwidth]
{
\subfigure[]{
\includegraphics[width=0.5\columnwidth]{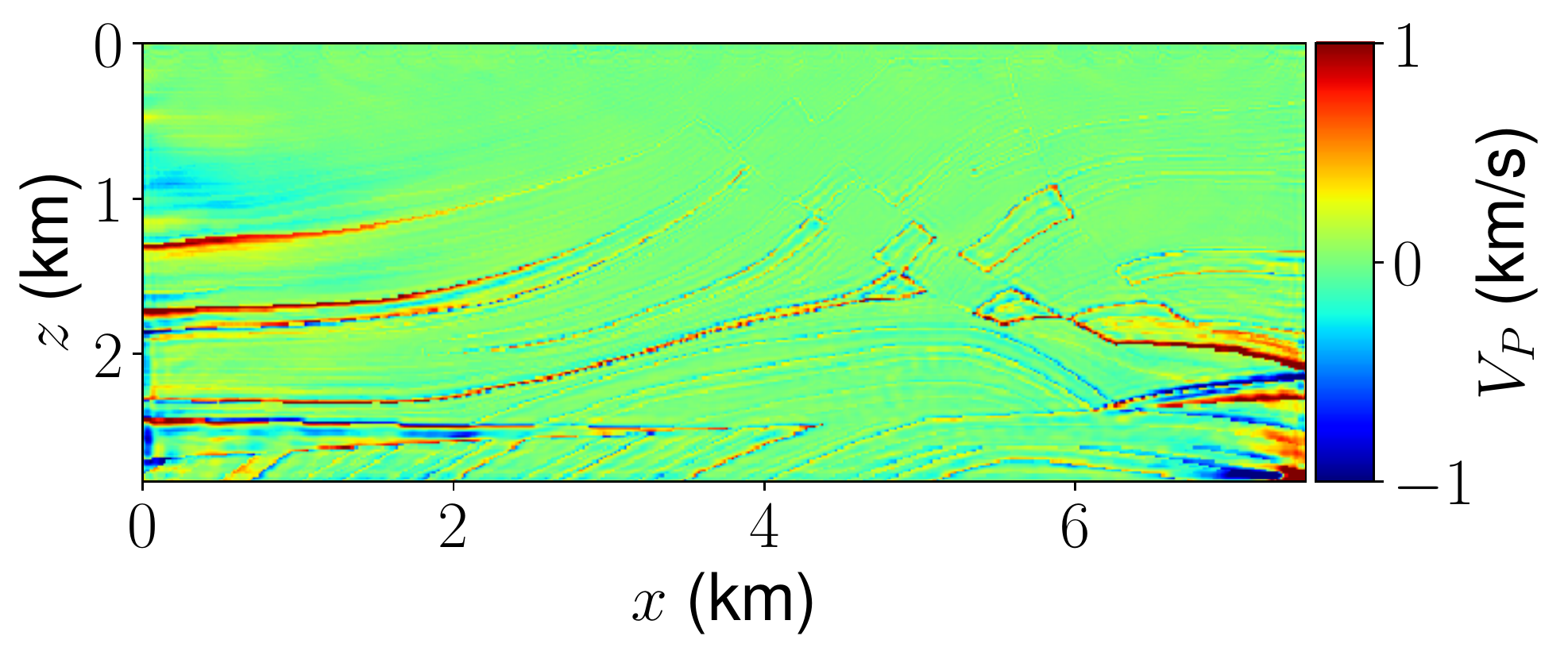}
\label{fig5-a}
}
\subfigure[]{
\includegraphics[width=0.5\columnwidth]{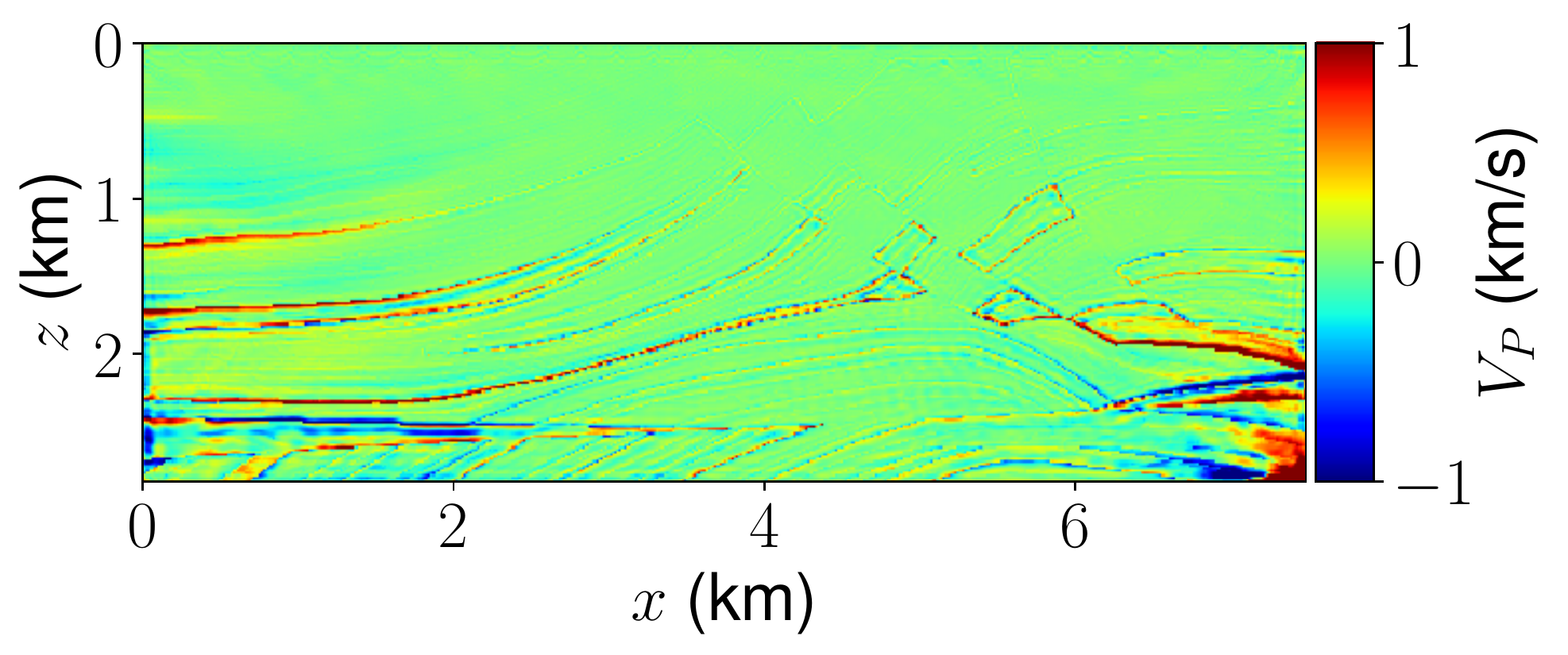}
\label{fig5-b}
}
}
\vspace{-0.1in}
\noindent\makebox[\columnwidth]
{
\subfigure[]{
\includegraphics[width=0.5\columnwidth]{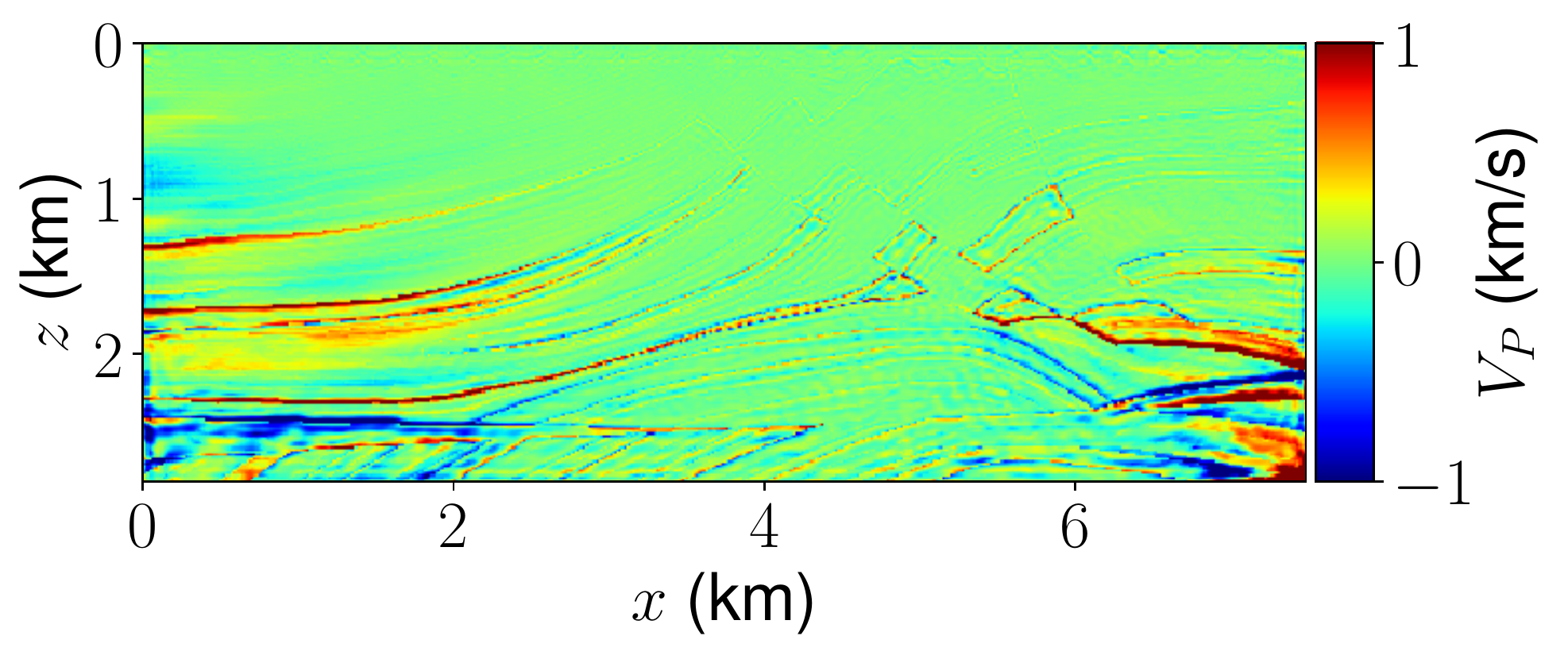}
\label{fig5-c}
}
\subfigure[]{
\includegraphics[width=0.5\columnwidth]{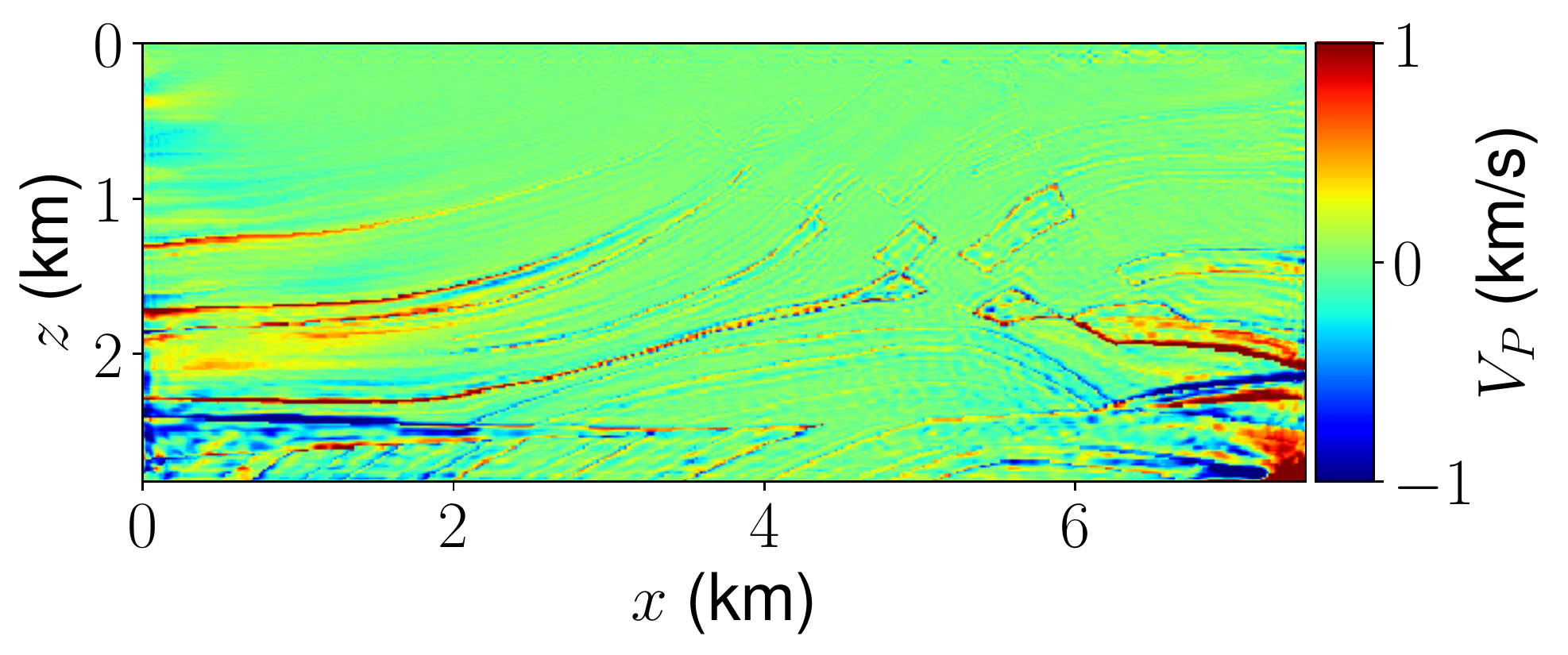}
\label{fig5-d}
}
}
\vspace{-0.1in}
\caption
{The difference between the true and retrieved velocity model after 64 epochs and a mini-batch size of (a) 4,  (b) 8, (c) 16. The velocity perturbation model in (d) is retrieved after 128 epochs with the full-batch of data (32 shots). }
\label{fig5}
\end{figure}

Figures~\ref{fig4}a, b, and c display the velocity model retrieved after 64 epochs on mini-batches with batch sizes of 4, 8, and 16, respectively. The full-batch with 32 shots required  128 epochs to retrieve  a velocity model similar to that of inversion with mini-batches (Figure~\ref{fig4}d). The difference between the true and inverted velocity models are plotted in Figure~\ref{fig5}.  While  all the inverted models are sufficiently close to the true models,  the velocity model corresponding to a mini-batch size of 8 exhibits superior resolution and is closest to the true model. The largest deviations are observed close to the edges of the model due to low data foldage.   The numerical examples demonstrate that mini-batch FWI with Adam optimization algorithm can reduce the computational cost and yield accurate results.

\section*{Discussion}

The numerical experiments outlined in the previous section illustrate the efficacy of the Adam optimization algorithm in  FWI with random shot selection.  We  found (results not shown here) that FWI with l-BFGS  method with the full-batch of 32 shots required a larger number of iterations to converge than the Adam optimization algorithm. Moreover, the velocity model was less resolved and required simulated data from around 100 shots to produce results with the resolution comparable to the Adam algorithm. Second-order methods are not regularly employed in training deep neural networks as they have been found to terminate near saddle points \citep{deeplearningbook}. However, accounting for the Hessian matrix is crucial in FWI to resolve  trade-off between different parameters encountered in for instance, elastic FWI \citep{FWI_overview2}. We plan to test the suitability of first order stochastic gradient-descent algorithms like Adam in multi-parameter FWI problems.

\section*{Conclusions}

We presented a strategy for full waveform inversion with random shot selection based on Adam, a first order stochastic gradient descent algorithm. The algorithm builds the search direction for model updates based on exponentially weighted  first- and second-order  moment estimates of the gradients from successive iterations. We outline an empirical strategy to select the hyperparameters of the Adam algorithm, particularly the learning rate, and the optimal mini-batch size for random shot selection. The algorithm is tested on synthetic data from the Marmousi model. We found that full waveform inversion with random shot selection can yield accurate results while leading to savings in computational time and resources.

\newpage
\onecolumn

\bibliographystyle{seg}  
\bibliography{example}

\end{document}